\documentclass[11pt]{article}
\usepackage{moriond2000,epsfig}

\def\Journal#1#2#3#4{{#1} {\bf #2}, #3 (#4)}

\def \aa   {\em Astr. \& Astrophys.}

\def \apj  {\em Astrophys. J.}

\def \nat  {\em Nature}

\font \bolditalics = cmmib10
\def \vc #1{{\textfont1=\bolditalics \hbox{$\bf#1$}}}

\def\pg{{\bf p}}

\def\eg{{\bf e}}

\def\thetag{{\vc \theta}}

\def\gammag{{\vc \gamma}}

\def\be{\begin{equation}}
\def\ee{\end{equation}}
\def\bea{\begin{eqnarray}}
\def\eea{\end{eqnarray}}

\begin{document}
\vspace*{4cm}
\title{COSMIC SHEAR WITH THE VLT}

\author{ R. MAOLI$^{1,2}$, Y. MELLIER$^2$, L. VAN WAERBEKE$^{2,3}$, 
P. SCHNEIDER$^{4,5}$, \\
B. JAIN$^6$, T. ERBEN$^5$, F. BERNARDEAU$^{7}$, B. FORT$^2$ }

\address{$^1$ Dipartimento di Fisica, Universit\`a di ``Roma La Sapienza'', Italy\\
$^2$ Institut d'Astrophysique de Paris, France\\
$^3$ Canadian Institute for Theoretical Astrophysics, Toronto, Canada\\
$^4$ Institut f\"ur Astrophysik und Extraterrestrische Forschung, Bonn, Germany\\
$^5$ Max Planck Institut f\"ur Astrophysik, Garching, Germany\\
$^6$ Department of Physics, Johns Hopkins University, Baltimore, USA\\
$^7$ Service de Physique Th\'eorique, C.E. de Saclay, France}

\maketitle\abstracts{
We report on the detection of cosmic shear on angular scales of 1.3-6.5 arcmin 
using 45 independent empty fields observed with the Very Large Telescope 
(VLT). This result confirms previous measurements obtained with the CFH 
Telescope at the same angular scales. We present the data analysis and the 
first preliminary results.
}

\section{Introduction}
The large scale structures of the Universe induce 
  gravitational  distortion of the shape of 
the background galaxies which can be observed on CCD images as a 
correlated ellipticity distribution of the lensed sources 
  (the cosmic shear). 
The analysis of cosmic shear permits to extract information on the geometry of
 space ($\Omega$ and $\lambda$) as well as on the power spectrum of the
(dark) matter 
density perturbations responsible for the distortion.
 Therefore, it directly  compares observations 
with cosmological models, without regards on the light distribution.

In contrast to weak lensing analysis of clusters of galaxies, 
  measuring cosmic shear is still a challenge. It has a very small amplitude ($\approx$ 5 times lower than 
 in clusters)  and demands  deep images of a 
large portion of the sky
obtained with high precision instruments. Up to now, this task was
beyond the reach of the available
technology. Although a  first detection in a blank sky region was 
reported two years ago~\cite{SvM98}, it is only recently that four 
different groups have succeeded in observing cosmic 
shear~\cite{vME00}$^,$~\cite{BRE00}$^,$~\cite{WTK00}$^,$~\cite{KWL00}.

The critical issues regarding the reliability of the cosmic shear measurement
are the three sources of noise:\\
\hspace*{2cm}- intrinsic noise\\
\hspace*{2cm}- systematics\\
\hspace*{2cm}- cosmic variance\\
Minimizing the first one is important to get a significant detection; understanding and 
controlling the second one is important 
to be sure that the signal is induced by  cosmic shear; then minimizing 
the third one is important to compare the signal with cosmological scenarios.

By using jointly these detections it is possible to rule out some scenarios.
 However,  the cosmic variance is still too large to distinguish among the 
more common models of the Universe.
A significant improvement implies the observation of many independent
fields.
The Cambridge group~\cite{BRE00} used nearly statistical independent 
areas of the sky, but has only 13 fields. The other experiments 
cover a larger area but sample compact regions with few independent fields.
In this case  the amplitude of the cosmic variance is estimated by 
running numerical simulations for each current cosmological models. 
In any case,  the cosmic variance is 
greater than in the case of observation of completely statistical independent
fields.  The alternative is to observe many independent fields in order
to infer the cosmic variance from the fluctuations of the shear observed
from field-to-field.  This is 
the strategy we developed with the VLT/FORS1 survey. In the following,
we describe the analysis of  cosmic shear in 50 FORS1 fields.
This is a preliminary study
which does not include an extensive investigation of systematics. Nevertheless,
the signal is strong and highly significant, so we are confident in the
results shown in this proceeding. 

\section{Observational strategy  analysis of VLT/FORS1 data}
All the data were obtained in the I band with the FORS1 instrument, 
a 2048$\times$2048 pixel CCD with a $6.8'$ field of view, mounted at the Cassegrain 
focus of the VLT-UT1.

\begin{figure}[t]
\center\psfig{figure=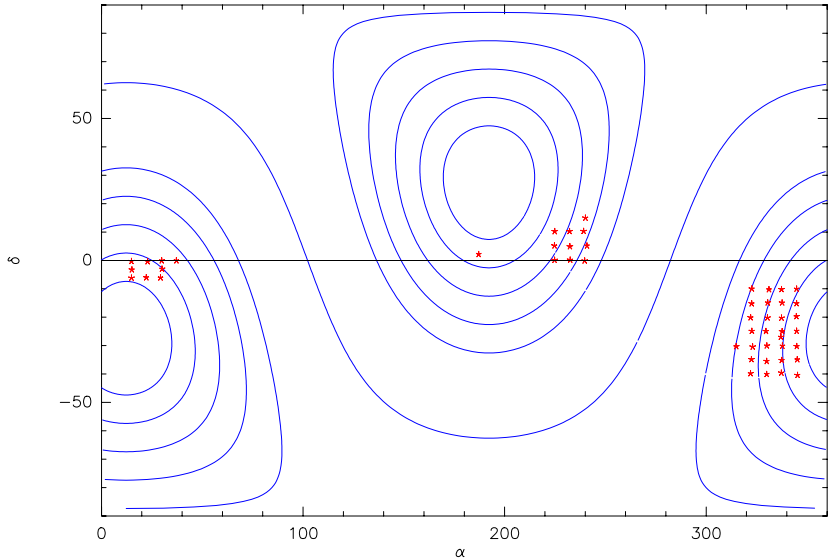,height=12.0cm,angle=270}
\caption{Distribution on the sky of the 50 VLT fields. Continuous lines are
for galactic latitudes $b=0,\pm 30,\pm 40,\pm 50,\pm 60,\pm 70$. Contrary 
to what it can seem to a first look, they are not on a regular grid.
\label{fig:skymap}}
\end{figure}

In figure~\ref{fig:skymap} we plot the sky distribution of the 50 VLT 
fields. Their main characteristics are:\\
\hspace*{0.5cm}- at least 5 degrees of separation, to be statistically 
independent;\\
\hspace*{0.5cm}- absence of bright stars ($m_B > 14$), to avoid diffraction 
spikes and light scattering;\\
\hspace*{0.5cm}- intermediate galactic latitudes 
($30^{\circ} \le b \le 70^{\circ}$), to have enough stars for the PSF 
correction;\\
\hspace*{0.5cm}- two fields in common with the HST/STIS project and two 
with the WHT project, to allow for cross checking of the results.

Observations were made between May and September 1999 using the service mode 
available at ESO. This observing strategy is particularly well suited for
cosmic shear surveys, ensuring a seeing always lower than 0.8'' and taking 
advantage from the possibility to spread the targets over a very large part 
of the sky. 

\begin{figure}[t]
\center\psfig{figure=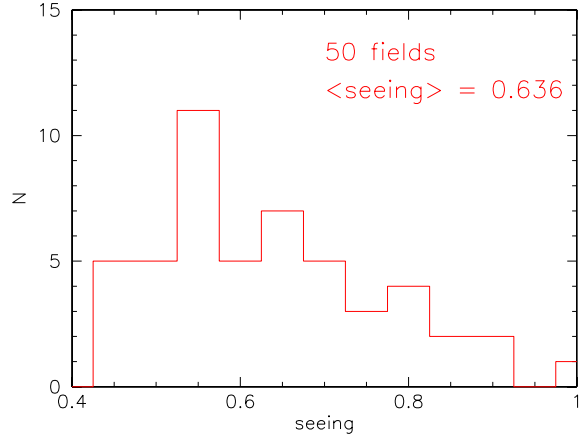,height=8.0cm,angle=270}
\caption{Histogram of the seeing for the 50 final coadded fields.
\label{fig:seeing}}
\end{figure}

Each field was observed for 36 minutes with 6 individual exposures per 
field distributed over a circle of $10^{\prime\prime}$ of radius. 
The median seeing for all the fields is $0.64^{\prime\prime}$ (see 
figure~\ref{fig:seeing}) 
and the limiting magnitude is $I_{AB}\sim 24.5-25$.

\begin{figure}[t]
\center\psfig{figure=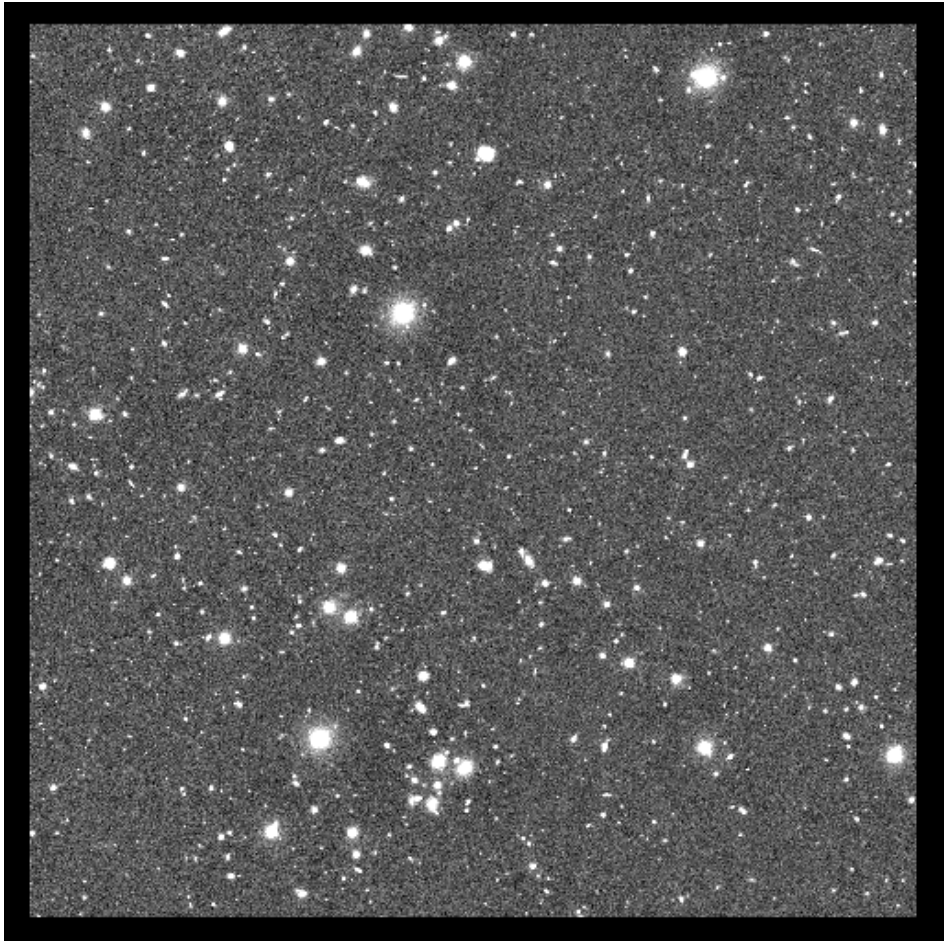,width=13.0cm}
\caption{Final coadded image of a VLT field. In this case the seeing is 
$0.53^{\prime\prime}$, the source density is $35.3 \ {\rm sources/arcmin^2}$,
with a $6.5^\prime\times 6.5^\prime$ field and with 67 not saturated stars.
\label{fig:image}}
\end{figure}

Each image was overscan corrected, bias subtracted and flatfielded with a 
superflat computed with the exposures of the same night. No fringe correction 
was necessary.
The FORS1 images get through a four port readout, so we treated separately 
the image subsets coming from the four quadrants of the CCD.
Then we normalized the gain of the different subsets to obtain an image with 
an homogeneous background. Finally, we aligned the six individual exposures
to produce the final coadded image. An example is given in 
figure~\ref{fig:image}.

At this stage, the images are ready for weak lensing analysis.
The incoming light, crossing the atmosphere and 
getting through the optical system of the telescope, undergoes a smearing 
and a circularization. 
If we define the ellipticity of the source as 
\be
\eg=\left({I_{11}-I_{22}\over Tr(I)} ; {2I_{12}\over Tr(I)}\right) \ \ \ 
{\rm with} \ \ I_{ij}=\int {\rm d}^2\theta\, W(\theta)\,\theta_i\theta_j\, 
f(\thetag) ,
\label{eq:ell}
\ee
the observed ellipticity can be expressed by the KSB 
formalism~\cite{KSB95}$^,$~\cite{LK97}$^,$~\cite{Hoe98},
\be
\eg^{obs}=\eg^{source}+P_\gamma\gammag+P^{sm} \pg
\label{eq:eobs}
\ee
where $\eg^{source}$ is the intrinsic ellipticity of the source, 
$P^{sm}$ is the smear polarizability tensor associated with the effects 
produced by the anisotropic part of the point spread function (PSF from now 
on) and $\pg$ is connected to the anisotropic kernel $g(\thetag)$ of the PSF 
through the two equations:
\be
\pg = (q_{11}-q_{22} ; 2 q_{12}) ; \ \ \ q_{lm} = \int {\rm d}^2\theta\, 
\theta_l\theta_m\, g(\thetag) .
\label{eq:PSFanis}
\ee

The vector $\pg$ can be computed from the stellar ellipticity $\eg^\star$
\be
p_\alpha={e_\alpha^\star\over P^{sm}_{\alpha\alpha}}.
\label{eq:estar}
\ee

The quantity $P_\gamma$ is the preseeing shear polarizability tensor, that 
takes into account both the effect of the gravitational shear and the 
circularization effect of the isotropic part of the PSF. It is given by the equation
\be
P^\gamma=P^{sh}-{P^{sh}_\star\over P^{sm}_\star} P^{sm}
\label{eq:Pgamma}
\ee
where $P^{sh}$ is the shear polarizability tensor and the subscript
``$\star$'' refers to the quantity computed for stars.

Assuming that there is no preferred direction for the intrinsic ellipticity 
of the sources, the cosmic shear at a given angular scale is 
\be
\gamma=\left<P_\gamma\right>^{-1}\cdot\left<\eg^{obs}-P^{sm} \pg\right>
\label{eq:gamma}
\ee
where the average is computed over a given angular scale.

Two main tasks can be identified in shear computation:\\
\hspace*{1cm}- separation between stars and galaxies and selection of stars
to compute the PSF correction for the galaxies;\\
\hspace*{1cm}- selection of galaxies for which the ellipticity can be computed
with a small error.

Practically, this is translated in the following steps
 (for details, see~\cite{vME00}).\\
a) {\it Masking of images:} we remove all the boundaries of the four quadrants 
of the CCD. this includes the external boundaries and the central cross 
delineating the four quadrants. We remove also all sources too close to 
bright sources that can deform them, satellite traces and in 
general all those sources whose shape is difficult to compute correctly.
About 12\% of objects are removed in this way.\\
b) {\it Stars selection:} as usual, we use a radius vs magnitude plot to 
select not saturated stars. A low cut in the magnitude or an high cut in 
the radius would result in selecting small galaxies instead of 
stars therefore producing lower quality PSF corrections.\\
c) {\it Fit of the stellar polarizability tensors:} using stars, we compute 
$p_\alpha={e_\alpha^\star\over P^{sm}_{\alpha\alpha}}$ and 
${P^{sh}_\star\over P^{sm}_\star}$ in all the image, fitting with a third 
order polynomial.\\
d) {\it $P^\gamma$ computation:} using the stellar fits, we compute 
Eq.~\ref{eq:Pgamma} for all the sources. $P^\gamma$ is a very noisy matrix 
with huge fluctuations. For a given source, we then compute $P^\gamma$ making 
an average among its values for the 30 nearest neighbors (see van Waerbeke
{\it et al.}, 2000).\\
e) {\it Source selection:} we select sources with stellar class 
(given by SExtractor) $<\, 0.9$, with radius $>$ stellar radius 
and with separation $>$ 10 pixels.

\begin{figure}[t]
\center\psfig{figure=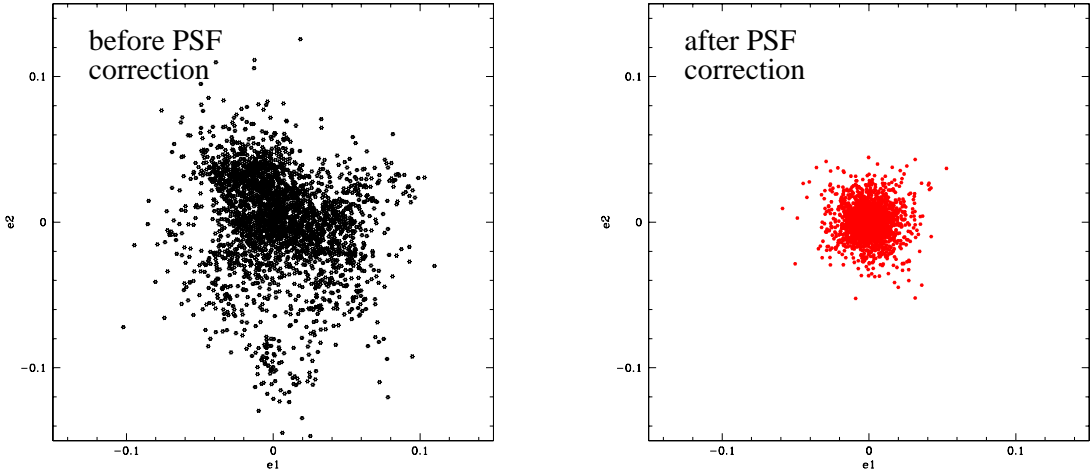,angle=270,width=15.0cm}
\caption{Star ellipticities before (at right) and after (at left) the PSF
correction, for all the 45 selected VLT fields. A good PSF correction implies 
a zero mean value for $e_1$ and $e_2$ with small symmetric fluctuations.
\label{fig:PSF}}
\end{figure}

In figure~\ref{fig:PSF} we plot the star ellipticities for all VLT fields 
before and after the PSF correction. Five fields had a poor PSF correction
and were eliminated.

At the end we have 45 fields with in average 60 stars/field (a minimum of 20 
and a maximum of 120); we have roughly 58700 sources for 
1900 ${\rm arcmin^2}$ that is $\sim 31 \ {\rm sources/arcmin^2}$ and 1300
sources/field.

\section{Results and conclusion}
For each final corrected image, we computed the variance of the shear,
$\left<\gamma^2\right>$, at different angular scales, where the angular scale
is given by the angular dimension of the top-hat window function.
Then we made an average of $\left<\gamma^2\right>$ over the 45 fields.

In this way we can associate an error to the value of $\left<\gamma^2\right>$
at a given scale. It is important to stress that with 45 fields this also
includes the cosmic variance.

\begin{figure}[t]
\center\psfig{figure=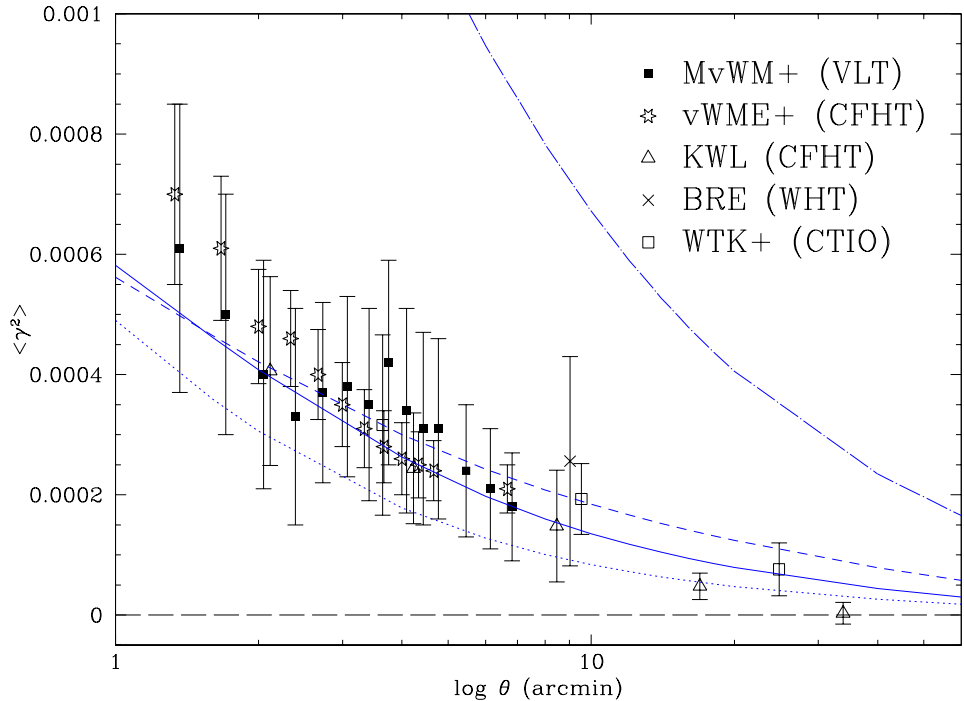,angle=270,width=13.4cm}
\caption{Summary of cosmic shear measurements. This work is referred as
MvWM+. Some predictions of current models are also plotted, assuming 
sources $z_{eff}=1$. The solid line corresponds to $\lambda$CDM, 
with $\Omega_m=0.3$, $\lambda=0.7$, $\Gamma=0.21$; the dot-dashed line
corresponds to COBE-normalized SCDM; the dashed line corresponds to 
cluster-normalized SCDM and the dotted line corresponds to 
cluster-normalized Open CDM with $\Omega_m=0.3$. 
\label{fig:results}}
\end{figure}

In figure~\ref{fig:results} we plot our results together with those of the
other cosmic shear measurement projects. Note the remarkable agreement between
the VLT and the CFHT results, although they were obtained with 
 different observational strategies. This is the best 
guarantee that cosmic shear measurement is not associated with any local
systematic effect.

Different theoretical models are also plotted in the figure. 
This plot rules out COBE-normalized standard CDM, which show the 
important potential of cosmic shear for cosmology. However, the 
errors are still too large to give precise predictions on the value of 
cosmological parameters $\Omega_0$ and $\Lambda$ 
 and the shape of the power spectrum of primordial density
fluctuations. Only a larger amount of data together with a complete control
of the systematics will allow to reach this important goal.

\section*{Acknowledgments}
We thank the ESO service observing team at the VLT for performing the 
observations. We thank V. Charmandaris (Dr IRAF), D. Bacon and 
A. Refregier for helpful discussions.
This work was supported by the TMR Network ``Gravitational Lensing: New
Constraints on Cosmology and the Distribution of Dark Matter'' of the EC 
under contract No. ERBFMRX-CT97-0172.


\section*{References}
\begin{thebibliography}{99}
\bibitem{SvM98}P. Schneider {\it et al}, \Journal{\aa}{333}{767}{1998}.

\bibitem{vME00}L. van Waerbeke {\it et al}, \Journal{\aa}{358}{30}{2000}.

\bibitem{BRE00}D. Bacon, A. Refregier \& R. Ellis, {\em MNRAS submitted}, 
astro-ph/0003008.

\bibitem{WTK00}D. Wittman {\it et al}, \Journal{\nat}{405}{143}{2000}.

\bibitem{KWL00}N. Kaiser, G. Wilson \& G. Luppino, 
{\em Astrophys. J. submitted}, astro-ph/0003338.

\bibitem{KSB95}N. Kaiser, G. Squires \& T. Broadhurst, 
\Journal{\apj}{449}{460}{1995}.

\bibitem{LK97}G. Luppino \& N. Kaiser, \Journal{\apj}{475}{20L}{1997}.

\bibitem{Hoe98}H. Hoekstra {\it et al}, \Journal{\apj}{504}{636}{1998}.

%
\end{thebibliography}
\end{document}